# Hardware/Software Security Patches for Internet of Trillions of Things

**John A. Stankovic, Tu Le, Abdeltawab Hendawi, Yuan Tian**
*University of Virginia*

*With the rapid development of the Internet of Things, there are many interacting devices and applications. One crucial challenge is how to provide security. Our proposal for a new direction is to create "smart buttons" and collections of them called "smart blankets" as hardware/software security patches rather than software-only patches.*

## 1.0 Introduction

The Internet of Things (IoT) market is forecasted to reach 30.7 billion devices in 2020 and grow to 75.4 billion devices by 2025.[1] At such a rapid pace, more and more new connected devices, as well as applications such as autonomous vehicles, smart homes, and smart cities, are appearing on the Internet. In the future, the Internet of Trillions of Things (IoTT) will include trillions of things (smart devices), with many different types of devices and communications, and an increasing variety of applications. Applications will both, directly and indirectly, interact with users, devices, other applications, and the environment. For example, in smart home environments, there already exists IoT devices such as security cameras, smart locks, garage door openers, smart thermostats, smoke detectors and they are all connected on the Internet. There are also Industrial IoT systems where perhaps thousands of devices are monitoring and controlling a process control plant. These IoT-based industrial plants may interact with other plants with just-in-time delivery of natural resources required by the plants, with humans in the plants, and with decision makers. Most industries (if not all) will make use of smart IoT devices.[2] Smart cities are seeing an ever-increasing deployment of sensors, actuators, and services to monitor and control transportation, emergency services, pollution, energy, health, etc. Once large IoT systems are installed, they will potentially exist for a long time.

For the IoT, security is extremely challenging, in general, and will be even more challenging in the future when IoTT comes into play. New challenges for security in IoTT are due to many factors. The interactions with the physical world through these trillions of smart devices have increased the attack surface dramatically. Attacks generally span across 3 layers: physical, communication, and application. For the physical layer, the fact that the smart devices exist in open environments exposes them to tampering. Attackers can also attack the devices indirectly by changing the surrounding environment that the devices are monitoring. Communications are mostly via various wireless technologies, allowing easy access to devices and applications. The great heterogeneity among devices and communications complicates security solutions. Applications and services will often include software that is installed on the device to control it. Applications and services can also include communications between devices via software that is installed on the system to manage the workflow of many connected devices.

As a new direction and because of the new security problems raised by the IoTT, we introduce a new type of security patch that combines hardware and software (HW/SW). The goal for this device called a "smart button" is to improve response to security attacks for IoTT, especially because of the physical aspects of these systems. Collections of smart buttons will become "smart blankets" which can be considered a type of security HW/SW middleware targeted towards security protection. Smart buttons will operate at 3 levels of focus: the physical, communication, and application layers in the IoTT. For each layer, we can create specific parameterized buttons to form a basic repository of HW/SW security patches. When attacks are detected, the right sets of smart buttons are selected, configured, and deployed as smart blankets. When there is a large number of attacks happening at the same time, solutions will include blankets of smart blankets. The repository of HW/SW security patches will grow over time as the attack-solution arms race continues. It is important to note that each IoTT system will have its own security measures (e.g., redundancy, secure keys, encryption, blockchains). We are not proposing ideas to build a secure IoTT application in the first place, but to "cover" it with our smart buttons and blankets when unanticipated attacks occur and **when standard software-only patching techniques fail to work**, or it is not feasible to replace devices and reinstall the software with the new fixes.

## 2.0 What is a HW/SW patch, its benefits, and challenges?

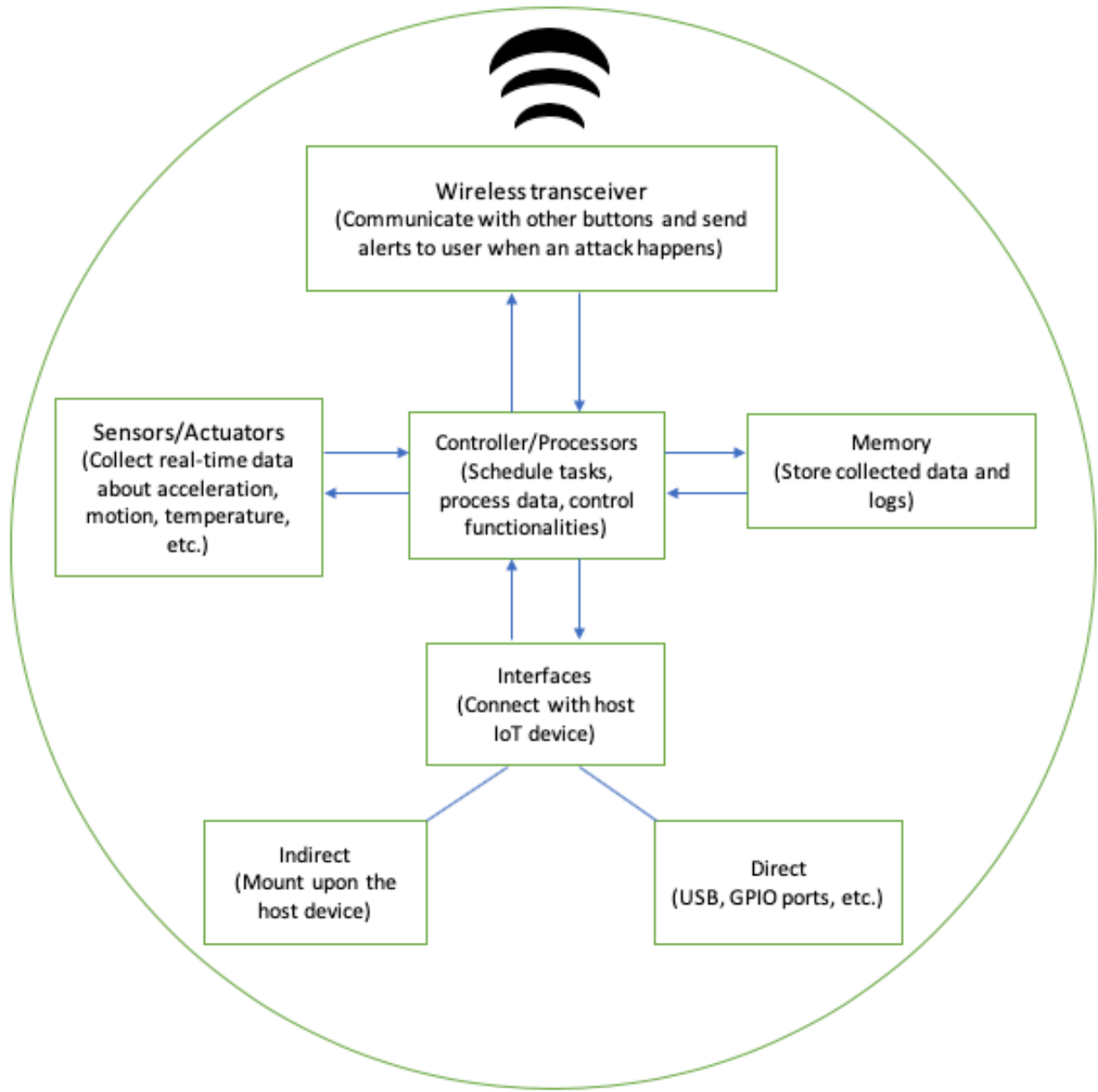

Figure 1: Overview structure of a smart button. Collections of these constitute

Smart Blankets

    A HW/SW patch (a smart button - See Figure 1) combines sensors, actuators, processors, and memory with associated software to protect IoT devices against one or more specific security attacks. In some cases, it also includes wireless communications to report the attack. It is meant to be used in IoTT systems where the physical world and smart devices are key components. When collections of HW/SW patches are deployed together in a cooperative fashion, this is called a smart blanket.

The smart buttons and blankets proposed solutions have several advantages:
- They can be applied to legacy devices and legacy applications/services as billions of them already exist and it is not feasible to retrofit or exchange them in a short term.
- Concepts found in the smart buttons can be useful for future designs of secure smart devices. In other words, IoTT devices should themselves become more supportive of security over time and this smart button solution is a mechanism to learn what they should look like.
- Even devices with previous protections embedded in them may not be able to handle zero-day attacks (i.e., new attacks that have never been seen before). In this case, new buttons with new defense mechanisms can then be added.
- In some situations, smart buttons can be used for extra security BEFORE any attacks are detected. For example, disaster response teams may be planning for deployment due to an environmental disaster and install buttons to help thwart an adversary who might try to exacerbate the environmental disaster situation.
- We can create a smart button repository to facilitate quickly creating HW/SW patches when they are needed for a specific attack.
- Because our smart button solution is a HW/SW patch (i.e., a device), it directly targets the physical nature of IoTT.

Despite the potential advantages, there are many open research challenges for smart buttons and blankets. These challenges include:
- The time and effort to create smart buttons.
- The deployment strategy, energy issues, and maintenance for smart buttons.
- How to protect the buttons themselves?
- What attack models can this solution address?
- What can individual buttons do themselves? When and how is the cooperation between the buttons required?
- How to handle the scale of trillions of things on the Internet?
- How to interface the buttons to IoTT systems?

## 3.0 Case Studies

This section demonstrates the value of smart buttons and blankets through hypothetical case studies. The case studies, in total, show the value of our approach for physical, communication, and application layers of an IoTT system. It is important to note that we have developed prototypes of HW/SW solutions for three out of these four case studies.

To better illustrate the core concepts, interfacing issues are not discussed in these examples. However, interfacing (See Figure 1) is a critical challenge and we devote Section 4 to interfacing discussion. Another critical challenge is secondary attacks (i.e., attacking the smart buttons after they are deployed). These case studies assume that there is no attack on the buttons themselves. However, we address this issue in Section 5.

### *3.1 Physical Layer Attack:*
**Consider a fixed-position surveillance camera.** Assume that a smart camera was installed in the front door of a house, and it is supposed to be mostly remain fixed in position. In this case, any movement of the camera that is not made by the legitimate user such as physically repointing or relocating the camera is considered malicious. To detect this attack, it may be possible to install a software-only patch on the camera that relies on the camera's capabilities such as image and video capture of background and detecting that the background is moving to detect attacks. However, leveraging such capabilities requires a lot of computational power that the camera may not have enough resources to provide or the internals of the camera software may be proprietary. In this scenario, if a smart button is used (e.g., adding an accelerometer and software to detect movement via the accelerometer and send an alarm if necessary), there are no changes to the camera and this attack can be detected. We built a prototype of the smart button to address this attack scenario. Our button provides acceleration sensing capability from an accelerometer, which the camera does not have in the first place, to detect any malicious movement of the camera.

Note that there are many other attack situations where a similar solution can apply, but perhaps other sensors are used such as temperature, vibration, gyroscopes, etc. As one example, if an attacker has infiltrated the control software for keeping the temperature of a chemical process below a threshold, that attacker can cause overheating. A new HW/SW patch that is independent of the previous system can be added to avoid this attack!

### *3.2 Communication Layer Attack:*

**Consider jamming.** Since the IoTT will rely heavily on wireless communications, addressing security attacks in this area are paramount. Jamming is a popular type of denial-of-service attack technique that leverages the transmission of signals to block the target wireless communications. Jamming attacks can vary in mechanisms and implementations. These attacks can exist in any IoTT application domain, e.g., smart homes and smart cities. Consider an interrupt jamming attack that is simple to perpetrate by using an attacking node (i.e., jammer) that disrupts communication. In interrupt jamming, the jammer only transmits when there are valid radio activities, which ensures stealthy attacks over a long term as signal transmission is energy-consuming. In previous work, we developed DEEJAM[3] as a novel MAC-layer protocol for defeating stealthy jammers with IEEE 802.15.4-based hardware. This solution has four layered HW/SW defensive mechanisms to hide the jammed system communication from a jammer and to communicate with a new channel. This solution is a HW/SW solution and could be considered as one concrete example of a smart button for a set of jamming attacks. The performance of the protocol was shown to be robust against successively more complex attacks: interrupt jamming addressed by frame masking, activity jamming addressed by channel hopping, scan jamming addressed by packet fragmentation, and pulse jamming addressed by redundant encoding. In summary, if a jamming attack is detected, we can add a smart button with DEEJAM protocol to each node in the system and protect against a set of jamming attacks on the original system.

### *3.3 Application Layer Attack:*

**Consider Stuxnet worm attack.** The Stuxnet worm attack on Iranian Nuclear Facilities discovered in 2010[4] is a very famous attack targeting programmable logic controllers (PLC). Basically, when the Stuxnet worm infects a PLC, it looks for the software that controls devices (e.g., manufacturing devices, robot arms, power plant controllers, etc.) and patches the software with malicious code. Stuxnet has functionality to hide itself, send malicious commands to the devices controlled by PLCs, and send fake feedback to users as if everything is working correctly. However, it cannot hide the physical actions of the devices it controls. In this case, smart buttons with independent hardware capabilities can more easily detect the Stuxnet attack rather than the software patch alone on the device. Further, any new software-only security patch once installed may itself be subject to the Stuxnet logic. In general, depending on the smart devices involved in the original system, the HW/SW patch might use motion, temperature, magnetic, electric field, etc. sensing.

### *3.4 Cross-Layer Attack:*

**Consider Amazon Echo.** Currently, many users are using virtual personal assistant (VPA) services such as Amazon Echo or Google Home in their home to control other smart home devices. However, these VPAs are vulnerable to hidden voice commands which are recognized by devices but not by humans.[5] For example, an attacker can leverage some attack vectors in smart home environment that play sounds such as television, radio station, and smartphones in order to send obfuscated voice commands embedded in streaming videos or podcasts to control a user's Amazon Echo. In this case, one can use a smart button to differentiate the legitimate user's voice command from the remote attacker's voice command. If the legitimate user sends a critical command to the Echo device in their living room, the smart button will send an audio pulse followed by post-processing to determine if that user is actually present in the room. In particular, the device can analyze the reflections received by a microphone array to sense presence of the person. In previous work, we built a prototype of this solution for the Amazon Echo and achieved 93.13% accuracy of identifying attacks during critical commands.[6]

## 4.0 Interfacing Issues

A key research challenge is interfacing the smart buttons to the IoTT and to each other. The challenges vary depending on the situation. We consider 5 categories of interfaces: indirect, direct with little or no assumptions about the application, direct with knowledge of the application, button to button, and blanket to blanket. In each case, there are two aspects that the button must address: detection and action. Note that smart buttons are expected to have their own energy sources so we do not discuss any interfacing issues with using the energy of the original system, even though sometimes it may be possible to do that.

**Indirect.** In some situations, a button will be mounted upon a "smart thing/device" and need only monitor it in indirect ways. For example, we can attach a smart button to a smart camera that has fixed orientation and the smart button can monitor if the camera is moved (via an accelerometer in the smart button). In this case, no direct interaction is required for the device. Once a violation occurs, the smart button must wirelessly report it as an alarm to a monitoring station. Such a solution can apply in many cases that protect against unwarranted movement, change in speed of movement, excess vibrations, an increase in temperature, etc.

**Direct without much knowledge.** Since a button is a HW/SW entity, it can attach to IoTT devices and applications in various ways. For situations where the button designer is not aware of the details of devices and/or applications, we propose to treat the smart button as an input/output device with a driver. The driver may be specific to a given IoT device and/or an operating system, so this abstraction allows buttons to connect to any existing device with a known interface with minimal new development beyond the driver that bridges the button application programming interface (API) to the existing interface.

**Direct with knowledge.** In some situations, the smart buttons deployment team might be aware of the internals of the smart IoT devices. For example, a processor of a smart IoT system can execute as a fog computing processor. That means that the fog device may be running various big data analytics and attempting to protect its local data with security-aware data structures. If the smart button is aware of these implementation details and can directly access the data (via a direct with knowledge interface), then more sophisticated security detection and actions will be possible. For instance, the smart button may detect that the data structure is corrupted or has received bad values and whether that the analytics program is not executing properly.

**Button to button.** Since smart buttons are created by the security team, their interface to each other (via the wireless transceivers) can be standard. There would be a button-to-button communication interface with the following characteristics: wirelessly communicates, can reach at least 100 feet to potentially limit the numbers of buttons needed (e.g., LoRa can achieve this distance), encrypts data and control signals during communication, can utilize different frequencies, can detect jamming, and supports transferring (exchanging) of security intrusion detection, attack information, and blanket control commands. The button-to-button communication network can be used directly as a redundancy method for detecting or protecting communication layer attacks to the IoTT network.

**Blanket to blanket.** Our hypothesis is that once we add a smart blanket to a system we can consider it as an integral part of the original system, so adding yet another blanket should be no different than adding the first one, and it can follow the interfacing strategies described above. However, various optimizations may be possible across blankets. For example, a more sophisticated feature in the second blanket may obviate the need for one or more protections in the first blanket.

## 5.0 Subsequent Attacks: Attacking the Security Buttons

Once a smart button or smart blanket is deployed to protect against a particular security attack, the attacker can attempt to attack the HW/SW patches. If the attack is successful, then either the previous smart blankets must be replaced with new smart buttons that protect against both the original system attack and the new attack, or one can consider a new smart blanket of HW/SW patches to be added on to the previous solution - a type of layers of smart buttons.

Note that collections of smart buttons forming the smart security blanket for a given security attack will communicate among themselves and offer various redundancies and diversity modalities. For example, the buttons can communicate with each other with security properties similar to those discussed above for protecting the communications of IoTT services. They can form consensus on both sensor readings and control actuations by the redundancy they provide. They can also use alternative sensing (orthogonal) modalities to avoid various physical attacks. There are many consensus schemes that can be implemented. Blockchain is an open, decentralized digital ledger technology that creates a secure way for the exchange of data. It is being widely touted and applied to IoT at the application layer. It might be possible to develop a lightweight blockchain for smart blankets. Hashing, proof of-work, and consensus found in blockchains may be too costly in time, memory, and energy requirements to be an effective smart blanket mechanism.

However, the size of a blanket and its lifetime are mitigating factors that can be exploited in attempting to develop a lightweight blockchain solution. If blockchain proves too costly, one might consider less expensive consensus solutions (albeit with less protection).

## 6.0 Discussion

The open IoTT environment where systems of systems are dynamically interacting in direct and indirect ways produces increasingly complicated attack surfaces. Software patches alone may often work. However, they are constrained by existing hardware capabilities of the device. Furthermore, sophisticated attacks always have mitigation techniques against defense mechanism (i.e., trying to hide from defense systems on the target device or disable them). Thus, successfully defending against those types of attacks may require installing additional hardware capabilities to the device. Installing security patches either in software form or hardware form on a system often requires shutting it down or reboot it as we can see in existing computing platforms. For example, installing Microsoft Windows updates always asking (if not forcing) the users to restart their computers to apply changes. Another example is when people need to send the product back to the manufacturer for repair or replacement due to hardware security issues. Security solutions must operate over long lifetimes where the dynamic evolution of the environment and apps are continuously happening. For example, smart cities transportation may deploy 10,000 smart devices for vehicles which must last for many years and evolve as the city evolves. Often we cannot shut down such a large running system to install security patches due to the fact that it takes a lot of time to restart large systems. Also, these systems can include critical operations which may cause severe problems if they are not working for even a short period of time. Imagine if the traffic control system is shut down for a moment, there would be a lot of traffic jams and accidents. A consequence of these complexities for IoTT is that the security attacks arms race will get worse and new approaches are required. We also hypothesize that there is no way to prevent all security attacks a priori. The system of systems is too dynamic, uncertain, heterogeneous, and continuously evolving. The security arms race will not end soon and a HW/SW patch is another tool that can sometimes be a better solution than traditional patching techniques.

Given that many IoTT applications will be large-scale, it is necessary to consider developing solutions that do not require EVERY device (sensor/actuator) to have a button attached to it. For example, if there are 10,000 smart devices in the subway system of New York City and there is a new attack requiring smart buttons, we need to develop solutions that can protect the system WITHOUT attaching a button to EVERY one of the 10,000 devices. In this case, new solutions must be created where individual buttons can protect sets of devices. The approach is twofold. One, smart buttons can act as HW/SW patch hubs. The hubs (analogous to routing hubs) will aggregate information from sets of devices, detect, and protect the devices based on the cumulative properties of those devices. Two, smart buttons can act as part of the fog, typically executing on a more powerful machine to provide local processing.

## 7.0 Related Work

IoT security area is growing in importance. Due to the rapid development of IoT devices, there exists a set of emergent threats to smart homes. New capabilities of smart home technologies lead to new consequences of some traditional attacks and enable more sophisticated attacks on IoT devices, including illegal physical entry and privacy violations.[7] There has been a lot of research on security and safety risks of IoT in three different categories: hardware, communication, and application.

**Hardware.** Recently, extensive research is going on to explore new security attacks on IoT devices. Researchers presented vulnerabilities of existing smart locks that would allow attackers to gain sensitive user information and unauthorized access to the home.[8] They highlighted design flaws, diversity in smart devices with different types of sensors, and lack of proper administration as key factors in those attacks. Other studies have shown that proper protections against malicious attacks are missing for sensors largely used in the smart devices. Sensors are vulnerable to spoofing by transduction attacks. Even without special-purpose equipments, an adversary can easily exploit physics to manipulate the outputs of sensors, causing unexpected behaviors of the systems that rely on those sensors. For example, the attacker can play sounds embedded in a Youtube video to control the output of a smartphone's microelectromechanical system accelerometer.[9]

**Communication.** IoT devices will be using fog computing which acts as an intermediate layer for securing the data stored in the cloud. However, the integration of IoT devices with the public cloud introduces additional security and safety threats.[10] A well-practiced architecture for smart home is to connect multiple IoT devices to a single smart hub or router. As IoT devices can connect to a WiFi network for communicating with the cloud, it becomes easier to connect them all to the same network and enable communications between them using that local network. Unfortunately, neither a WiFi router nor the protocol used in the router was initially designed to support this type of connectivity among the smart devices, as a result, it poses a new security challenge. The presence of a security firewall (i.e., preventing illegal access of one device from the other devices) in the network layer is emerging as multiple devices are operating on the same network and some of them might get compromised due to the lack of security. And with the help of the malicious device, an attacker can attack other devices connected to the same network, causing massive damage to the smart home users. In this case, the firewall would serve as a defense, protecting all other devices from the infected one. Researchers discussed a security manager platform built on top of an IoT hub to monitor usage patterns of all IoT devices in the home and detect anomalous network activities.[11] Furthermore, for short-range and power-efficient communication purpose especially in smart home environment, ZigBee and Z-Wave protocols are widely used in IoT devices. Researchers identified the existence of threats in those protocols due to implementation failures and shortfalls. A practical security analysis showed that Zigbee protocol was designed for easy setup and usage, thus exposing vulnerabilities that would allow attackers to easily jam the communication or sniff the transmitted network key.[12] An implementation error in Z-Wave key exchange protocol could allow attackers to take full control of a door lock by using a low-cost Z-Wave packet interception and injection tool.[13] In some cases, the misuse of some network protocols can cause severe security and privacy problems. People largely believe in the security of autonomous system and they often consider it as a trustworthy system.

**Application.** There has been some recent research on security analysis of smart home applications. Researchers analyzed Samsung's SmartThings which has the largest number of applications among currently available smart home platforms and presented an attack where malicious applications can steal sensitive information such as lock codes of the IoT devices.[14] Their analysis showed that over 55% of applications in SmartThings marketplace are overprivileged. Hence, an application can gain full access to a device even if it only needs limited access to the device. Although IoT platforms such as SmartThings require users to grant permissions to applications, malicious applications tend to request for unneeded permissions and users often do not know what those applications actually do behind the scenes. Additionally, the interaction chains between applications can pose high risks to smart home users. A recent study on 185 official SmartThings applications has shown that 37 out of 162 hidden inter-application interaction chains through physical surroundings are considered highly dangerous to the safety of users.[15]

## 8.0 Conclusion

In this article, we propose a novel research direction to handle security attacks in the era of IoTT. This proposed solution approach is based on adding integrated hardware and software patches as a security monitoring and protection layer to the existing devices (things). One important lesson learned from this work is that software-only patches can solve many security issues in IoTT, but not all of them. Many security issues cannot be detected by software-only patches due to missing hardware capabilities in the device. Thus, we introduce the smart button and smart blanket solution to fill these gaps. Although being an effective solution, smart buttons and blankets still have many open challenges such as protecting themselves from attacks, synergistic interaction between different types of buttons in a form of a blanket to protect against complex security scenarios, and smooth interfacing with existing IoT devices. With new solutions for these issues, security defenses for the future IoTT can be significantly improved.

**John A. Stankovic** is the BP America Professor in the Department of Computer Science at the University of Virginia. His research interests are in cyber physical systems, the Internet of Things, and Smart Health. His PhD is from Brown University. Contact him at stankovic@cs.virginia.edu.

**Tu Le** is a PhD student in the Department of Computer Science at the University of Virginia. His research interests are in Security & Privacy, Internet of Things, and Human-Computer Interaction. Contact him at tnl6wk@virginia.edu.

**Abdeltawab Hendawi** is a Research Associate in the Department of Computer Science at the University of Virginia. His research interests are in big data management and analytics, and smart cities. Contact him at hendawi@virginia.edu.

**Yuan Tian** is an Assistant Professor in the Department of Computer Science at the University of Virginia. Her research interests are in security and privacy, and the Internet of Things. Her PhD is from Carnegie Mellon University. Contact her at yuant@virginia.edu.